\documentclass[AMA,STIX1COL]{WileyNJD-v2}

\articletype{Research Article}%

\received{1 December 2020}
\revised{N/A}
\accepted{N/A}

\usepackage{graphicx}
\usepackage{mathrsfs}
\usepackage{amssymb}
\usepackage{amsfonts}
\usepackage{amsmath}
\usepackage{epsfig}
\usepackage{color}
\usepackage{epstopdf}
\usepackage{algorithm}
\usepackage{algpseudocode} 
\usepackage{multirow,array}
\usepackage{setspace}
\usepackage{algorithm}
\usepackage{algpseudocode} 
\usepackage[T1]{fontenc}
\usepackage[utf8]{inputenc}
\usepackage{mathtools}

\begin{document}

\title{Wireless Secret Sharing Game between Two Legitimate Users and an Eavesdropper}

\author[1]{Lei Miao*}

\author[2]{Dingde Jiang}

\authormark{Lei Miao \textsc{et al}}

\address[1]{\orgdiv{Mechatronics Engineering}, \orgname{Middle Tennessee State University}, \orgaddress{\state{Tennessee}, \country{USA}}}

\address[2]{\orgdiv{School of Astronautics \& Aeronautic}, \orgname{University of Electronic Science and Technology of China}, \orgaddress{\state{Sichuan}, \country{China}}}

\corres{*Lei Miao, MTSU, Box 19, Murfreesboro, TN 37132, USA. \email{lei.miao@mtsu.edu}}

\abstract[Summary]{Wireless secret sharing is crucial to information security in the era of Internet of Things. One method is to utilize the effect of the randomness of the wireless channel in the data link layer to generate the common secret between two legitimate users Alice and Bob. This paper studies this secret sharing mechanism from the perspective of game theory. In particular, we formulate a non-cooperative zero-sum game between the legitimate users and an eavesdropper Eve. In a symmetrical game where Eve has the same probability of successfully receiving a packet from Alice and Bob when the transmission distance is the same, we show that both pure and mixed strategy Nash equilibria exist. In an asymmetric game where Eve has different probabilities of successfully receiving a packet from Alice and Bob, a pure strategy may not exist; in this case, we show how a mixed strategy Nash equilibrium can be found.}


\maketitle

\section{Introduction}\label{sec1}

Security and privacy in wireless networking relies on symmetric-key cryptography, which requires pre-established private keys at both the transmitter and the receiver. In the era of Internet of Things (IoT) where Machine to Machine (M2M) communications frequently occur with minimum human intervention, automatic and secure sharing of secrets for the purpose of cryptography is crucial to information security. There are various ways to share secrets automatically in wireless networks. One direction is to combine cryptographic schemes and channel coding techniques so that transmitted messages between two legitimate users Alice and Bob cannot be decoded by the eavesdropper Eve \cite{BarrosRodrigues} \cite{Maurer93}. Recent works along this line can be found in \cite{chen2019secure}, \cite{hyadi2020securing}, and \cite{mukherjee2018secure} for interference, broadcast, and  multiple access channels, respectively. It has also been studied on using cooperation such as cooperative relaying and jamming \cite{wang2015enhancing} to enhance wireless secrecy. Some recent applications of physical layer security can be found in \cite{wang2016physical} and \cite{zhang2016secrecy} for 5G systems. Another approach exploits the principle of reciprocity \cite{Balanis_Antenna} in wireless communications and extracts the secret from the common observation between Alice and Bob on the wireless channel state \cite{MathurTrappe08} \cite{MiaoDiffSecretSharing} \cite{ruotsalainen2019experimental}. All these methods mentioned above are collectively known as the physical layer solutions, which essentially exploit the randomness and varying nature of wireless channels to share secrets. They do not work very well when the speed of variation in wireless channels is slow and may also require costly modifications to existing communication protocols and infrastructure.

In a different direction, the effect of wireless channel dynamics on the data link layer is utilized to share secrets \cite{XiaoGongTowsley2010}, \cite{YaoFukui09}, \cite{safaka2012creating}. The idea behind works along this line is as follows: Alice and Bob keep sending each other unicast packets without retry, using which the secret is derived; Eve would eventually lose a packet and be unable to figure out the secret even if she knew exactly the mechanism Alice and Bob use. More details of this approach can be found in our previous work \cite{miao2019optimal} where we discuss optimal secret sharing between Alice and Bob with the presence of Eve. In particular, we assume in \cite{miao2019optimal} that Eve's location is random, and only Alice and Bob can choose how to generate the secret; we show that when the probability of successfully transmitting a packet is monotonically decreasing with the transmission distance and Eve's location is uniformly distributed, the optimal strategy for Alice and Bob to minimize the probability that Eve figures out the secret is to generate half of the secret from each one of them.

In this paper, we consider the case that Eve can also choose her location in order to maximize her probability of receiving all packets and figuring out the secret. Specifically, we assume that both the legitimate users (Alice and Bob) and the eavesdropper (Eve) do not know each other's strategy but are both rational. Let $P_e$ be the probability of Eve figuring out the secret. Then, Alice and Bob's goal is to minimize $P_e$ or maximize $-P_e$, and Eve's goal is to maximize $P_e$. This observation motivates us to formulate the problem as a zero-sum game between the legitimate users and the eavesdropper. 

Security games have been studied extensively on the interaction between legitimate and malicious users, and game-theoretic approaches have been applied to a wide range of problems, including security at the physical and MAC layers,  security at the application layer, cryptography, etc. For comprehensive reviews, see \cite{manshaei2013game} \cite{abdalzaher2016game} \cite{do2017game}. Our secret sharing game is different from the existing ones in the literature: we study how to share secrets using the effect of the unreliable nature of wireless channels on the data link layer. Our results are based on the probability function of Eve successfully receiving a packet. Nonetheless, our analysis does not rely on a specific form of the probability function; instead, our work would be applicable to any probability function as long as a mild assumption is satisfied. The main contributions of this paper is as follows: \textit{(i)} We show that the optimal secret sharing problem can be considered as a game between two legitimate users and the eavesdropper; \textit{(ii)} We analyze the symmetric game case and identify both pure and mixed strategy Nash equilibria; \textit{(iii)} For the asymmetric game case, we discover two different scenarios that yield pure and mixed Nash equilibrium, respectively; and \textit{(iv)} We show how the mixed strategy Nash equilibrium can be found when the probabilities of successful packet transmission are known.  

The organization of the rest of the paper is as follows: in Section \ref{sec2}, we discuss the system model and formulate the game; in Section \ref{sec3}, we present the main results of the optimal secret sharing zero-sum game; and finally, we conclude in Section \ref{section4}. Due to space constraint, all the proofs are omitted but are available in \cite{security_gametheory_arxiv}.

\section{System Model and Problem Formulation}\label{sec2}

The summary of system symbols and their definitions can be found in Table \ref{symbol_definition}. In our system model, the two legitimate users Alice and Bob are at two different locations that are $D$ meters away, and they are trying to exchange $N$ packets $\{Pkt_1, Pkt_2, \dots, Pkt_N\}$, using which the secret is calculated. See Fig. 3 in \cite{miao2019optimal} for illustration for a packet exchange process. One simply way to obtain the secret is to exclusive-OR all $N$ packets together: $secret=Pkt_1 \oplus Pkt_2 \oplus \dots \oplus Pkt_N$. Due to the unreliable nature of wireless communications, Eve will have high probability of losing one or more packets when $N$ is large so that she will not be able to figure out the secret. Without loss of generality, we let $N$ be an even number. For ease of notation, We assume that each of the two game players, i.e., the legitimate users and the eavesdropper, has three strategies. For Alice and Bob, there are totally $N+1$ strategies, which can be represented by: (1) $S_{A,n}$: Alice sends $N-n$, $n\in\{0,1,\dots,\frac{N}{2}-1\}$, packets to Bob, and Bob sends $n$ packets to Alice; (2) $S_{B,n}$: Bob sends $N-n$, $n\in\{0,1,\dots,\frac{N}{2}-1\}$, packets to Alice, and Alice sends $n$ packets to Bob; and (3) $S_{AB}$: each one of them sends $N/2$ packets to the other. Eve chooses to stay somewhere between Alice and Bob, and she has three different strategies: stay close to Alice, stay close to  Bob, and stay in the exact middle. We use $L_A$, $L_B$, and $L_M$ to denote these three locations/strategies, respectively. We further assume that locations $L_A$ and $L_B$ are $\epsilon$, $\epsilon \in (0,\frac{D}{2})$, meter away from Alice and Bob, respectively; location $L_M$ is $\frac{D}{2}$ meters away from both Alice and Bob. Thus, $P_A(\epsilon)$, $P_A(D-\epsilon)$, and $P_A(\frac{D}{2})$ are the probabilities of Eve successfully receiving a packet from Alice when Eve's strategy is $L_A$, $L_B$, and $L_M$, respectively. Similarly, $P_B(\epsilon)$, $P_B(D-\epsilon)$, $P_B(\frac{D}{2})$ are the probabilities of Eve successfully receiving a packet from Bob when Eve's strategy is $L_B$, $L_A$, and $L_M$, respectively. 

Let $P_A(d)$ and $P_B(d)$ be the probability of Eve successfully receiving a packet from Alice or Bob, respectively, when the transmission distance is $d$. We have the following assumption about $P_A(d)$ and $P_B(d)$.

\begin{assumption}
	\label{assumption1}\textit{(i)} Each packet transmission is independent from each other; \textit{(ii)} $P_A(d)$ and $P_B(d)$ are time-invariant; \textit{(iii)} $P_A(\epsilon)>P_A(\frac{D}{2})>P_A(D-\epsilon)$ and $P_B(\epsilon)>P_B(\frac{D}{2})>P_B(D-\epsilon)$; and  \textit{(iv)} $P_A(\frac{D}{2})>\frac{1}{2}[P_A(D-\epsilon)+P_A(\epsilon)]$ and $P_B(\frac{D}{2})>\frac{1}{2}[P_B(D-\epsilon)+P_B(\epsilon)]$.
\end{assumption}
The assumptions above is quite generic and does not require the exact form of functions $P_A(d)$ and $P_B(d)$. Parts \textit{(i)} and \textit{(ii)} above are valid in slow-fading environments where the coherence time of the wireless channel is long and the channel state is stable during the period of secret sharing. Part \textit{(iii)} states that the key factor that determines the probability of successful packet transmission is the distance, which is especially true in long-distance wireless communications. An example of $P_A(d)$ and $P_B(d)$  supporting the monotonicity assumption in VANET (Vehicular Ad Hoc Networks) environments can be found in \cite{killat2009empirical}, in which Killat et al. simulate and verify a theoretical probability of successful transmission function of distance inferred from the Nakagami-m distribution of RF wave propagation. It is well known that in free space, the path loss of RF signals is proportional to the square of distance. Part \textit{(iv)} above reflects this: in spite of random factors such as channel fading, the signal's power and the probability of successful transmission attenuates faster when the distance is larger; there is evidence in the literature showing that the probability is a concave function of distance, especially for short-distance wireless communications (see Fig. 1 in \cite{killat2009empirical}, Fig. 12 in \cite{xu2019energy}, Fig. 9 and Fig. 10 in \cite{cordeiro2003interference}, Fig. 6b in \cite{elbatt2006cooperative}, and Fig. 10 in \cite{lau1992capture}). 

\section{Optimal Secret Sharing as a Zero-Sum Game}\label{sec3}

\begin{table}
	\caption{Summary of symbols and their definitions}
	\small
	\centering
	\begin{tabularx}{\textwidth}{|l|X|l|X|}
		\hline
		\textbf{Symbols} & \textbf{Definition} &\textbf{Symbols}  &\textbf{Definition}\\ 
		\hline 
		$s_E$ & The strategy of Eve. &$L_A, L_B, L_M$ &  The possible strategies of Eve: stay close to Alice, stay close to Bob, and stay in the middle. \\ 
		\hline 
		$s_L$ & The strategy of the legitimate users (Alice and Bob). & $S_{A,n}, S_{B,n},S_{AB}$ & The possible strategies of Alice an Bob: Alice sends more packets, Bob sends more packets, and each sends exactly half. \\
		\hline 
		$P_A(d), P_B(d)$ & The success probability of Eve receiving a packet from Alice and Bob, respectively, as a function of distance. & $D, \epsilon$ & D: the distance between Alice and Bob. $\epsilon$: a small distance less than $D/2$.\\
		\hline
	\end{tabularx}%
	\label{symbol_definition}
\end{table} 

Let $s_{L}$ and $s_{E}$ be the strategies of the legitimate users, i.e., Alice and Bob, and Eve, the eavesdropper, respectively. We have $s_L\in\{S_{A,n}, S_{B,n}, S_{AB}\}$ and $s_E\in\{L_A, L_B, L_{M}\}$ We use $U_L(s_L,s_E)=-P_e$ and $U_E(s_L,s_E)=P_e$ to denote the utility functions of the legitimate users and Eve, respectively. Essentially, Alice and Bob would like to minimize the probability of Eve figuring out the secret, and Eve would like to maximize the same probability. 

\begin{definition}
	\label{Definition1}
	A strategy profile $(s_L^*,s_E^*)$ is a \textit{Nash equilibrium} if $U_L(s_L^*,s_E^*)\ge U_L(s_L, s_E^*)$ for each feasible strategy $s_L$ and $U_E(s_L^*,s_E^*)\ge U_E(s_L^*, s_E)$ for each feasible strategy $s_E$.
\end{definition}

\subsection{Symmetric Game}
We first consider a symmetric game scenario that the following hold:
\begin{equation*}
		P_A(L_A)=P_B(L_B)=P(\epsilon), 	\text{ } P_A(L_B)=P_B(L_A)=P(D-\epsilon) ,  \text{ and } P_A(L_M)=P_B(L_M)=P(D/2).
\end{equation*}

We have the utility matrix shown in Table \ref{tab1} where the utility functions of Eve are positive and the ones of Alice and Bob are negative. Next, let us first introduce an auxiliary lemma.
\begin{center}
	\begin{table*}[htpb]%
		\caption{Utility matrix of the symmetric game. \label{tab1}}
		\centering
		\vspace*{-\baselineskip}
		\begin{tabular}{cc|c|c|c|}
		& \multicolumn{1}{c}{} & \multicolumn{3}{c}{Alice and Bob}\\
		& \multicolumn{1}{c}{} & \multicolumn{1}{c}{$S_{A,n}, q_{A,n}$}  & \multicolumn{1}{c}{$S_{AB}, q_{AB}$} & \multicolumn{1}{c}{$S_{B,n}, q_{B,n}$} \\\cline{3-5}
		\multirow{3}*{Eve}  & $L_A, p_1$ & $\pm P^{N-n}(\epsilon)P^n(D-\epsilon)$ & $\pm P^{\frac{N}{2}}(\epsilon)P^{\frac{N}{2}}(D-\epsilon)$ &$\pm P^n(\epsilon)P^{N-n}(D-\epsilon)$ \\\cline{3-5}
		& $L_M, p_2$ & $\pm P^N(\frac{D}{2})$ & $\pm P^N(\frac{D}{2})$ & $\pm P^N(\frac{D}{2})$\\\cline{3-5}
		& $L_B, 1-p_1-p_2$ & $\pm P^{N-n}(D-\epsilon)P^n(\epsilon)$ & $\pm P^{\frac{N}{2}}(\epsilon)P^{\frac{N}{2}}(D-\epsilon)$ & $\pm P^{N-n}(\epsilon)P^{n}(D-\epsilon)$\\\cline{3-5}
	\end{tabular}
	\end{table*}
\end{center}

\begin{lemma}
	\label{lemma_auxiliary}	$P^{\frac{N}{2}}(\epsilon)P^{\frac{N}{2}}(D-\epsilon)<P^N(\frac{D}{2})$.
\end{lemma}
\textbf{Proof}: Because  $P(\epsilon)\in(0,1)$, $P(D-\epsilon)\in(0,1)$, and $P(\frac{D}{2}) \in (0,1)$, we only need to show that $P(\epsilon)P(D-\epsilon)<P^2(\frac{D}{2})$. 
Because $\epsilon\in(0,\frac{D}{2})$, $D-\epsilon \neq \epsilon$. From part \textit{(iii)} of Assumption \ref{assumption1}, we have
\begin{equation*}
	[P(D-\epsilon)-P(\epsilon)]^2=P^2(D-\epsilon)+P^2(\epsilon)-2P(D-\epsilon)P(\epsilon)>0,
\end{equation*}
i.e.,  
\begin{equation}
	\label{squre_greater}\frac{1}{4}[P^2(D-\epsilon)+P^2(\epsilon)]>\frac{1}{2}[P(D-\epsilon)P(\epsilon)].
\end{equation}
From part \textit{(iv)} of Assumption \ref{assumption1}, we have
\begin{equation*}
	P^2(\frac{D}{2})=P^2(\frac{1}{2}(D-\epsilon)+\frac{1}{2}\epsilon)>[\frac{1}{2}P(D-\epsilon)+\frac{1}{2}P(\epsilon)]^2 =\frac{1}{4}[P^2(D-\epsilon)+P^2(\epsilon)]+\frac{1}{2}[P(D-\epsilon)P(\epsilon)]
\end{equation*}
Invoking (\ref{squre_greater}), we have $P^2(\frac{D}{2})>P(D-\epsilon)P(\epsilon) \text{ }.\blacksquare$

We are now ready to discuss the pure strategy result of the symmetric game. 
\begin{lemma}\label{lemma_symmetric_game}
	Strategy profile $(S_{AB},L_M)$ is a pure strategy Nash equilibrium. 
\end{lemma}
Proof: It can be seen from the utility matrix that $U_L(S_{AB}, L_M)=U_L(S_{A,n}, L_M)=U_L(S_{B,n}, L_M)=-P^N(\frac{D}{2})$. Invoking Lemma \ref{lemma_auxiliary}, we have
\begin{equation*}
	U_E(S_{AB},L_M)=P^N(\frac{D}{2})>P^{\frac{N}{2}}(\epsilon)P^{\frac{N}{2}}(D-\epsilon)=U_E(S_{AB},L_A)=U_E(S_{AB},L_B).
\end{equation*}

From Definition \ref{Definition1}, it follows that strategy profile $(S_{AB},L_M)$ is a pure strategy Nash equilibrium. $\blacksquare$

Lemma \ref{lemma_symmetric_game} indicates that in the pure strategy Nash equilibrium, Alice and Bob each generates half of the packets and Eve stays in the middle location $L_M$. We now turn our attention to a mixed strategy Nash equilibrium, in which Eve has probabilities $p_1$, $p_2$, and $p_3=1-p_1-p_2$ to use strategies $L_A$, $L_M$, and $L_B$, respectively; similarly, Alice and Bob have probabilities $q_{A,n}$, $q_{AB}$, and $q_{B,n}$, sum of which is $1$, to use strategies $S_{A,n}$, $S_{AB}$, and $S_{B,n}$, respectively. 

\begin{lemma}
	\label{lemma_mixed_eve}In a mixed strategy Nash equilibrium, Eve's strategy is to stay at $L_M$ with probability 1;  Alice and Bob should have positive probabilities on all strategies $S_{A,n}$, $S_{B,n}$, and $S_{AB}$ so that: 		
	\begin{equation}
		\label{mixed_alice_bob_1}\sum_{n=0}^{\frac{N}{2}-1}q_{A,n}P^{N-n}(\epsilon)P^n(D-\epsilon)+q_{AB}P^\frac{N}{2}(\epsilon)P^\frac{N}{2}(D-\epsilon)+\sum_{n=0}^{\frac{N}{2}-1}q_{B,n}P^n(\epsilon)P^{N-n}(D-\epsilon)<P^N(\frac{D}{2})
	\end{equation}
	and
	\begin{equation}
		\label{mixed_alice_bob_2}\sum_{n=0}^{\frac{N}{2}-1}q_{A,n}P^{N-n}(D-\epsilon)P^n(\epsilon)+q_{AB}P^\frac{N}{2}(D-\epsilon)P^\frac{N}{2}(\epsilon)+\sum_{n=0}^{\frac{N}{2}-1}q_{B,n}P^{N-n}(\epsilon)P^{n}(D-\epsilon)<P^N(\frac{D}{2})
	\end{equation}
\end{lemma}
\textbf{Proof:} Suppose that $0<q_{A,n}<1$, $0<q_{AB}<1$, and $0<q_{B,n}<1$, $\forall n\in\{0,1,\dots,\frac{N}{2}-1\}$. In a mixed strategy Nash equilibrium, we have:
\begin{align*}
	&-p_1P^{N-n}(\epsilon)P^n(D-\epsilon)-p_2P^N(\frac{D}{2})-p_3P^{N-n}(D-\epsilon)P^n(\epsilon)\\
	&=-p_1P^{\frac{N}{2}}(\epsilon)P^{\frac{N}{2}}(D-\epsilon)-p_2P^N(\frac{D}{2})-p_3P^{\frac{N}{2}}(\epsilon)P^{\frac{N}{2}}(D-\epsilon)\\
	&=-p_1P^n(\epsilon)P^{N-n}(D-\epsilon)-p_2P^N(\frac{D}{2})-p_3P^{N-n}(\epsilon)P^n(D-\epsilon)
\end{align*}		
Solving the above equations, we get $p_1=p_3=0$, and $p_2=1$. If it is the case in the mixed strategy Nash equilibrium, we must also have (\ref{mixed_alice_bob_1}) and (\ref{mixed_alice_bob_2}). 

Next, we verify that when (\ref{mixed_alice_bob_1}) and (\ref{mixed_alice_bob_2}) hold, $\exists$ $0<q_{A,n}<1$, $0<q_{AB}<1$, and $0<q_{B,n}<1$, $\forall n\in\{0,1,\dots,\frac{N}{2}-1\}$. Let $q_{A,n}=q_{B,n}$, and (\ref{mixed_alice_bob_1}) and (\ref{mixed_alice_bob_2}) become one inequality:
\begin{equation}
	\label{mixed_alice_bob}
\sum_{n=0}^{\frac{N}{2}-1}q_{A,n}[P^{N-n}(\epsilon)P^n(D-\epsilon)+P^n(\epsilon)P^{N-n}(D-\epsilon)]+q_{AB}P^\frac{N}{2}(\epsilon)P^\frac{N}{2}(D-\epsilon)<P^N(\frac{D}{2})=2\sum_{n=0}^{\frac{N}{2}-1}q_{A,n}P^N(\frac{D}{2})+q_{AB}P^N(\frac{D}{2})
\end{equation}
Invoking Lemma \ref{lemma_auxiliary}, we have 
$	q_{AB}P^\frac{N}{2}(\epsilon)P^\frac{N}{2}(D-\epsilon)<q_{AB}P^N(\frac{D}{2})$.
We now consider two cases. \\
\textit{Case 1:} $P^{N-n}(\epsilon)P^n(D-\epsilon)+P^n(\epsilon)P^{N-n}(D-\epsilon) \le 2P^N(\frac{D}{2})$, for all $n$. In this case, (\ref{mixed_alice_bob}) always holds as long as $q_{A,n}$ are nonzero probabilities. \\
\textit{Case 2:} $P^{N-n}(\epsilon)P^n(D-\epsilon)+P^n(\epsilon)P^{N-n}(D-\epsilon) > 2P^N(\frac{D}{2})$, for some $n$. In this case, we can always pick small enough positive $q_{A,n}$  values so that (\ref{mixed_alice_bob}) holds.
$\blacksquare$

\subsection{Asymmetric Game}
We now consider an asymmetric game scenario that $P_A(d)>P_B(d)$, i.e., when the transmission distance is the same, Eve has higher probability to successfully receive a packet from Alice than from Bob. For example, if Alice has higher transmission power than Bob or Bob is closer to a noise source, then the signal to noise ratio between Alice and Eve may be higher than that between Bob and Eve, causing the asymmetric game scenario described above.  We have the following utility matrix shown in Table \ref{tab2}.
\begin{center}
	\begin{table*}[htpb]
		\caption{Utility matrix of the asymmetric game. \label{tab2}}
		\centering
		\vspace*{-\baselineskip}
		\begin{tabular}{cc|c|c|c|}
		& \multicolumn{1}{c}{} & \multicolumn{3}{c}{Alice and Bob}\\
		& \multicolumn{1}{c}{} & \multicolumn{1}{c}{$S_{A,n}, q_{A,n}$}  & \multicolumn{1}{c}{$S_{AB}, q_{AB}$} & \multicolumn{1}{c}{$S_{B,n}, q_{B,n}$} \\\cline{3-5}
		\multirow{3}*{Eve}  & $L_A, p_1$ & $\pm P_A^{N-n}(\epsilon)P_B^n(D-\epsilon)$ & $\pm P_A^{\frac{N}{2}}(\epsilon)P_B^{\frac{N}{2}}(D-\epsilon)$ &$\pm P_B^{N-n}(D-\epsilon)P_A^n(\epsilon)$ \\\cline{3-5}
		& $L_M, p_2$ & $\pm P_A^{N-n}(\frac{D}{2})P_B^n(\frac{D}{2})$ & $\pm P_A^{\frac{N}{2}}(\frac{D}{2})P_B^{\frac{N}{2}}(\frac{D}{2})$ & $\pm P_B^{N-n}(\frac{D}{2})P_A^n(\frac{D}{2})$\\\cline{3-5}
		& $L_B, 1-p_1-p_2$ & $\pm P_A^{N-n}(D-\epsilon)P_B^n(\epsilon)$ & $\pm P_B^{\frac{N}{2}}(\epsilon)P_A^{\frac{N}{2}}(D-\epsilon)$ & $\pm P_B^{N-n}(\epsilon)P_A^n(D-\epsilon)$\\\cline{3-5}
	\end{tabular}
	\end{table*}
\end{center}

\begin{lemma}
	\label{lemma_asymmetric_pure}If $P_A(d)>P_B(d)$, and $P_B(\epsilon)\le P_A(D-\epsilon)$, then strategy profile $(S_{B,0}, L_B)$ is a pure strategy Nash equilibrium. 
\end{lemma}
Proof: When $n=0$, we invoke part \textit{(iii) }of Assumption \ref{assumption1} and get 
\begin{equation*}
	P_B^N(\epsilon)>P_B^N(\frac{D}{2})>P_B^N(D-\epsilon), \text{i.e.,}
\end{equation*}
\begin{equation}
	\label{UE_LB}U_E(S_{B,0},L_B)>U_E(S_{B,0},L_M)>U_E(S_{B,0},L_A).
\end{equation}
By assumption $P_B(\epsilon)\le P_A(D-\epsilon)$, we get
\begin{equation}
	\label{Pre_U_AB}-P_B^N(\epsilon)\ge -P_B^{N-n}(\epsilon)P_A^n(D-\epsilon), \forall n\in\{1,2,\dots,\frac{N}{2}-1\}
\end{equation}
and 
\begin{equation}
	\label{Pre_U_AB_2}-P_B^N(\epsilon)\ge -P_B^{\frac{N}{2}}(\epsilon)P_A^{\frac{N}{2}}(D-\epsilon) \ge -P_A^{N-n}(D-\epsilon)P_B^n(\epsilon), \forall n\in\{0,1,\dots,\frac{N}{2}-1\}
\end{equation}
Combining (\ref{Pre_U_AB}) and (\ref{Pre_U_AB_2}), we get:
\begin{equation}
	\label{U_AB}U_{L}(S_{B,0},L_B) \ge U_{L}(S_{AB},L_B) \ge U_{L}(S_{A,0},L_B).
\end{equation}
From (\ref{UE_LB}) and (\ref{U_AB}), it follows that strategy profile $(S_{B,0}, L_B)$ is a pure strategy Nash equilibrium. $\blacksquare$

The intuition behind Lemma \ref{lemma_asymmetric_pure} is that if $P_B(d)$ is so much less than $P_A(d)$ so that $P_B(\epsilon)\le P_A(D-\epsilon)$, then the best strategy of the legitimate users is to always let Bob send the packets; conversely, the best strategy of Eve is to stay close to Bob so that she could maximize the probability of receiving all packets.

\begin{lemma}\label{lemma_asymmetric_no_pure}
	If $P_A(d)>P_B(d)$ and $P_B(\epsilon)>P_A(D-\epsilon)$, then there is no pure strategy Nash equilibrium.
\end{lemma}
Proof: By assumption, we have $P_A(\epsilon)>P_B(\epsilon)>P_A(D-\epsilon)>P_B(D-\epsilon)$. 
We discuss the three rows of the utility matrix individually.\\
(1) In the first row, we have the largest utility function $U_L(S_{B,0},L_A)=-P_B^N(D-\epsilon)$, which is greater than all other ones in the same row. Now, let us look at the column of $U_L(S_{B,0},L_A)$. Due to part  \textit{(iii)} of Assumption \ref{assumption1}, we have $U_E(S_{B,0},L_A)=P_B^N(D-\epsilon)<U_E(S_{B,0},L_M)=P_B^N(D/2)$. Therefore, there is no pure strategy Nash equilibrium in the first row of the utility matrix. \\
(2) In the second row, we have the largest utility function $U_L(S_{B,0},L_M)=-P_B^N(D/2)$, which is greater than all other ones in the same row. Now, let us look at the column of $U_L(S_{B,0},L_M)$. Due to part  \textit{(iii)} of Assumption \ref{assumption1}, we have $U_E(S_{B,0},L_M)=P_B^N(D/2)<U_E(S_{B,0},L_B)=P_B^N(\epsilon)$. Therefore, there is no pure strategy Nash equilibrium in the second row of the utility matrix. \\
(3) In the third row, we have the largest utility function $U_L(S_{A,0},L_B)=-P_A^N(D-\epsilon)$, which is greater than all other ones in the same row. Now, let us look at the column of $U_L(S_{A,0},L_B)$. Due to part  \textit{(iii)} of Assumption \ref{assumption1}, we have $U_E(S_{A,0},L_B)=P_A^N(D-\epsilon)<U_E(S_{A,0},L_M)=P_A^N(D/2)<U_E(S_{A,0},L_A)=P_A^N(\epsilon)$. Therefore, there is no pure strategy Nash equilibrium in the third row of the utility matrix. $\blacksquare$

Lemma \ref{lemma_asymmetric_no_pure} shows that when $P_B(\epsilon)>P_A(D-\epsilon)$, i.e., $P_B(d)$ is not too much less than $P_A(d)$, no pure strategy Nash equilibrium exists. According to \cite{nash1951non}, at least one mixed strategy Nash equilibrium always exists in this case. In what follows, we present the procedure to find such a mixed strategy equilibrium. For notation use, we let Alice and Bob have only three strategies: $S_{A,0}, S_{AB}, and S_{B,0}$. The utility functions are:
\begin{align}
\label{mixed_Sa_1}-p_1P_A^N(\epsilon)-p_2P_A^N(\frac{D}{2})-p_3P_A^N(D-\epsilon)\tag{$q_1$}\\
\label{mixed_Sab_1}
-p_1P_A^{\frac{N}{2}}(\epsilon)P_B^{\frac{N}{2}}(D-\epsilon)-p_2P_A^{\frac{N}{2}}(\frac{D}{2})P_B^{\frac{N}{2}}(\frac{D}{2})
-p_3P_B^{\frac{N}{2}}(\epsilon)P_A^{\frac{N}{2}}(D-\epsilon)
\tag{$q_2$}\\
\label{mixed_Sb_1}-p_1P_B^N(D-\epsilon)-p_2P_B^N(\frac{D}{2})-p_3P_B^N(\epsilon)\tag{$q_3$}\\
\label{mixed_La_1}q_1P_A^N(\epsilon)+q_2P_A^{\frac{N}{2}}(\epsilon)P_B^{\frac{N}{2}}(D-\epsilon)+q_3P_B^N(D-\epsilon)\tag{$p_1$}\\
\label{mixed_Lm_1}q_1P_A^N(\frac{D}{2})+q_2P_A^{\frac{N}{2}}(\frac{D}{2})P_B^{\frac{N}{2}}(\frac{D}{2})+q_3P_B^N(\frac{D}{2})\tag{$p_2$}\\
\label{mixed_Lb_1}q_1 P_A^N(D-\epsilon)+q_2P_B^{\frac{N}{2}}(\epsilon)P_A^{\frac{N}{2}}(D-\epsilon)+q_3P_B^N(\epsilon)\tag{$p_3$}
\end{align}
where \eqref{mixed_Sa_1}, \eqref{mixed_Sab_1}, and \eqref{mixed_Sb_1} are the payoffs of the legitimate users when strategies $S_{A,0}$, $S_{AB}$, and $S_{B,0}$ are used, respectively; \eqref{mixed_La_1}, \eqref{mixed_Lm_1}, and \eqref{mixed_Lb_1} are the payoffs of Eve when strategies $L_A$, $L_M$, and $L_B$ are used, respectively. The procedure of finding the mixed strategy Nash equilibrium involves two steps: \textit{proposition} and \textit{verification}. In the first step, we make an assumption about either $\{p_1,p_2,p_3\}$ or $\{q_1,q_2,q_3\}$ and use the utilization functions to solve for the other set of probabilities. If the solution is feasible and we are able to use it in the second step to verify that the proposition provided in Step 1 is indeed true, the Nash equilibrium is found. The formal procedure can be found in Algorithm \ref{alg_mixed_strategy} where we only show the propositions about $\{p_1,p_2,p_3\}$; the pseudo code of making propositions about $\{q_1,q_2,q_3\}$ is very similar.
\begin{algorithm}[thpb]
\caption{Finding mixed strategy Nash equilibrium in an asymmetric game when $P_B(\epsilon)>P_A(D-\epsilon)$  }\label{alg_mixed_strategy}
\begin{algorithmic}[1]
	\State \textbf{Proposition: enumerate the following assumptions.}
	\State{$p_1,p_2,$ and $p_3$ are all non-zero probabilities; solve \eqref{mixed_La_1}=\eqref{mixed_Lm_1}=\eqref{mixed_Lb_1} for $q_1,q_2,$ $q_3$ and go to Verification.}
	\State{For any two probabilities $p'$ and $p'' \in \{p_1,p_2,p_3\}$,  assume they are non-zero and use $p'''$ to denote the remaining probability. Solve ($p'$)=($p''$)>($p'''$) for $q_1,q_2,q_3$ and go to Verification.}
	
	\State \textbf{Verification:}
	
	\If{the solution is infeasible}
	\State Continue to the next assumption
	\Else		
	\If{$q_1$,$q_2$, and $q_3$ are all positive}
	\State {Solve \eqref{mixed_Sa_1}=\eqref{mixed_Sab_1}=\eqref{mixed_Sb_1} for $p_1,p_2,p_3$.}
	\EndIf
	
	\If{two probabilities $q'$ and $q'' \in \{q_1,q_2,q_3\}$ are positive, and the remaining probability $q'''$ is $0$ }
	\State {Solve $(q')=(q'')>(q''')$ for $p_1,p_2,p_3$.}
	\EndIf
	
	\If{$q' \in \{q_1,q_2,q_3\}$ is $1$, and the other two probability $q''$ and $q'''$ are $0$}
	\State {Solve $(q')>(q'')$ and $(q')>(q''')$ for $p_1,p_2,p_3$.}
	\EndIf

	\If{the solution of $p_1,p_2,p_3$ matches with the proposition}
	\State {Nash equilibrium is found and exit}
	\Else
	\State{Continue to the next assumption}
	\EndIf
	\EndIf
	
\end{algorithmic}
\end{algorithm}

\subsection{Numerical Example}
In this subsection, we present a numerical example. For ease of calculation, we let $N=2$ and assume that Alice and Bob have only three strategies: $S_{A,0}, S_{AB}, and S_{B,0}$.. The probabilities are: $P_A(\epsilon)=0.99$, $P_A(\frac{D}{2})=0.94$,  $P_A(D-\epsilon)=0.80$, $P_B(\epsilon)=0.90$, $P_B(\frac{D}{2})=0.84$, and $P_B(D-\epsilon)=0.70$. 
Invoking Lemma \ref{lemma_asymmetric_no_pure}, there is no pure strategy Nash equilibrium. The mixed strategy utility functions corresponding to \eqref{mixed_Sa_1} through \eqref{mixed_Lb_1} are:
\begin{align}
	\label{mixed_Sa}-0.3401p_1-0.2364p_2-0.64\\
	\label{mixed_Sab}0.027p_1-0.0504p_2-0.72\\
	\label{mixed_Sb}0.32p_1+0.1044p_2-0.81\\
	\label{mixed_La}0.4901q_1+0.203q_2+0.49\\
	\label{mixed_Lm}0.178q_1+0.084q_2+0.7056\\
	\label{mixed_Lb}-0.17q_1-0.09q_2+0.81
\end{align}

We start out by assuming that $p_1\in(0,1)$, $p_2\in(0,1)$, and $1-p_1-p_2\in(0,1)$. Under this proposition, we have $(\ref{mixed_La})=(\ref{mixed_Lm})=(\ref{mixed_Lb})$, whose solution is $q_1=1.946$, $q_2=-3.292$, and $1-q_1-q_2=2.346$. This is infeasible, meaning that $p_1$, $p_2$, and $1-p_1-p_2$ cannot be all positive and less than 1. Next, we discuss three cases of $p_1$, $p_2$, and $1-p_1-p_2$. \\
Case 1: $p_1\in(0,1)$, $p_2\in(0,1)$, and $1-p_1-p_2=0$. It yields that $(\ref{mixed_La})=(\ref{mixed_Lm})>(\ref{mixed_Lb})$. There are two solutions that ensure $q_1$, $q_2$, and $1-q_1-q_2$ are not all positive and less than 1. Therefore, we have two subcases: \\
Case 1.1: $q_1=0.6908$, $q_2=0$, and $1-q_1-q_2=0.3092$. It implies that at equilibrium, we must have $(\ref{mixed_Sa})=(\ref{mixed_Sb})>(\ref{mixed_Sab})$, which has no solution between $0$ and $1$ for $p_1$ and $p_2$.\\
Case 1.2: $q_1=0.5469$, $q_2=0.4531$, and $1-q_1-q_2=0$. It implies that at equilibrium, we must have $(\ref{mixed_Sa})=(\ref{mixed_Sab})>(\ref{mixed_Sb})$, which has solutions of $p_1$ and $p_2$ so that $p_1+p_2\in(0,1)$. It implies that $1-p_1-p_2\in(0,1)$, which is impossible.\\
Case 2: $p_1\in(0,1)$, $p_2=0$, and $1-p_1-p_2\in(0,1)$. It yields that $(\ref{mixed_La})=(\ref{mixed_Lb})>(\ref{mixed_Lm})$. There are no solutions.\\
Case 3: $p_1=0$, $p_2\in(0,1)$, and $1-p_1-p_2\in(0,1)$. In this last case, we have $(\ref{mixed_Lm})=(\ref{mixed_Lb})>(\ref{mixed_La})$. The only feasible solution to it is $q_1=0$, $q_2=0.6$, and $1-q_1-q_2=0.4$. If this solution is also the one in equilibrium, we need to have $(\ref{mixed_Sab})=(\ref{mixed_Sb})>(\ref{mixed_Sa})$, which also has a feasible solution: $p_1=0$, $p_2=0.5814$, and $1-p_1-p_2=0.4186$. \\
It completes the numerical example, and the mixed Nash equilibrium is: $(p_1,p_2,1-p_1-p_2)=(0,0.5814,0.4186)$ and $(q_1,q_2,1-q_1-q_2)=(0,0.6,0.4).$

\section{Conclusions}\label{section4}%
We have studied the optimal secret sharing problem between two legitimate users (Alice and Bob) and an eavesdropper (Eve), formulated as a non-cooperative zero-sum game. In the symmetric game case, both pure and mixed strategy Nash equilibria exist. Our results indicate that regardless of the type of the equilibrium, Eve should always stay in the middle of Alice and Bob. In the pure strategy Nash equilibrium, the best strategy of Alice and Bob is to generate half of the packets from each one of them; in a mixed strategy Nash equilibrium, Alice and Bob could generate all the packets from one user only, but some inequalities involving the probabilities must hold. 

In the asymmetric game case that Eve has better chance to successfully receive packets from Alice than from Bob, we show that there are two scenarios: if it is very asymmetrical, then a pure strategy Nash equilibrium exists, in which Bob is the one who generates all the packets and Eve chooses to stay near Bob; o.w., a mixed strategy equilibrium exists and can be calculated.

\end{document}